\documentclass[pra,aps,twocolumn,showpacs]{revtex4}
\usepackage{psfig}
\usepackage{bm}

\begin{document}

\title{Linear optics quantum Toffoli and Fredkin gates}

\author{Jarom\'{\i}r Fiur\'{a}\v{s}ek} 
\affiliation{Department of Optics, Palack\'{y} University, 17. listopadu 50,
77200 Olomouc, Czech Republic}

\date{\today}

\begin{abstract}
We design  linear optics multiqubit quantum logic gates. We assume the traditional 
encoding of a qubit onto state of a single photon in two modes (e.g. spatial or
polarization). We suggest schemes allowing direct probabilistic realization of the fundamental Toffoli and Fredkin
gates without resorting to a sequence of single- and two-qubit gates. This yields
more compact schemes and potentially reduces the number of ancilla photons. 
The proposed setups involve passive linear optics, sources of auxiliary 
single photons or maximally entangled pairs of photons, and single-photon detectors. 
In particular, we propose an interferometric implementation of the  
Toffoli gate in the coincidence basis, which does not require any ancilla photons
and is experimentally feasible with current technology.

\end{abstract}


\maketitle

\section{Introduction}

Quantum information theory  \cite{Nielsen00} exploits the laws of quantum 
mechanics to devise novel methods of information processing and transmission  
that would be impossible or very hard to achieve classically. 
During recent years various protocols for quantum information processing were 
successfully demonstrated experimentally with  several different 
physical systems. Particular attention has been paid to optical
realizations where the quantum bits are encoded onto states of single photons.
Photons are ideal carriers of quantum information because they can be
distributed over long distances in low-loss optical fibers or in free space. 
While perfect for
quantum communication purposes, photons seemed to be less suitable for 
quantum computing because the lack of sufficiently strong optical nonlinearities
seemed to prevent the implementation of quantum gates  between  photons. 

The situation changed radically in 2001 when Knill, Laflamme, and Milburn (KLM)
published their landmark paper in which they showed that a scalable universal
quantum computation is possible with only single photon sources, passive linear
optical interferometers  and single
photon detectors \cite{Knill01}. The key insight of KLM is that the 
nonlinearity (such as a Kerr effect) can be simulated on a single-photon level 
using the above listed resources, conditioning on particular measurement 
outcomes of the detectors and applying appropriate feedback. The resulting 
linear optics quantum gates \cite{Kok05}
are generally only probabilistic but the probability of success could be 
in principle made arbitrary close to unity by exploiting off-line generated 
multi-photon entangled states and  quantum teleportation
\cite{Bennett93,Bouwmeester97,Boschi97}. 

The KLM paper stimulated a number of further  works suggesting 
alternative and improved constructions of the basic quantum C-NOT gate 
\cite{Ralph01,Pittman01,Knill02,Hofmann02b,Franson02,Zou02b} 
whose experimental demonstrations by several groups followed 
\cite{Pittman02,Pittman03,Brien03,Sanaka04,Gasparoni04,Zhao05,Langford05,Kiesel05,Okamoto05}. 
However,  despite these promising successes, extending this approach 
to more complex schemes involving higher number of photons currently 
appears to be a formidable experimental task because
the overhead in resources (in particular the number of ancilla photons)
required by the original KLM scheme is very high.

It is possible to combine the ideas of one-way quantum
computation \cite{Raussendorf01,Raussendorf03} and linear optics quantum computing
to  significantly reduce the resources required for the computation.
The techniques introduced by KLM could be used to generate a multiphoton 
cluster state which then serves as a resource for quantum computing which
proceeds by performing certain carefully chosen measurements on each photon from
the cluster \cite{Yoran03,Nielsen04,Browne05,Gilbert05}. First
proof-of-principle experimental demonstration of one-way quantum computation with
four-photon cluster state has been reported recently \cite{Walther05b}.

In this paper, we wish to address a different aspect of the quantum computing with linear
optics. Namely, we will be interested in the implementations of the fundamental
Toffoli and Fredkin gate which play an important role both in classical (reversible)
computing and in quantum computing and information processing \cite{Nielsen00}. 
In the universal quantum computer the multi-qubit gates are usually assumed to be
implemented as a sequence of single and two-qubit gates. However, this strategy
may not be optimal in the context of linear optics quantum computing, where
schemes tailored specifically for multiqubit gates may require less ancilla
photons or achieve higher probability of success than implementations relying on
a sequence of the single and two-qubit gates.

The rest of the paper is organized as follows. In Section II we will present 
a scheme for the $N$-qubit generalized Toffoli gate, which flips the state 
of the $N$th qubit in the computational basis if all the $N-1$ control qubits 
are in state $|\tilde{1}\rangle$, where tilde indicates the logical qubit states
throughout the paper to distinguish them from the Fock states $|n\rangle$. 
We will first design gate operating in the so-called coincidence basis 
\cite{Ralph02},  which has the advantage 
that this scheme does not require any ancilla photons. To make this  
gate non-destructive it is necessary to perform quantum non-demolition measurement 
of number of photons at the output of the gate, which could be done with linear 
optics, ancilla photons and photon-number resolving detectors \cite{Kok02}. 
Section III is devoted to the three-qubit Fredkin gate which is a controlled SWAP
gate, the states of the two target qubits are swapped if the control qubit 
is in state $|\tilde{1}\rangle$. Our scheme requires only six ancilla photons. 
Finally, Section IV contains brief conclusions and summary of the main results.

\begin{figure}
\centerline{\psfig{figure=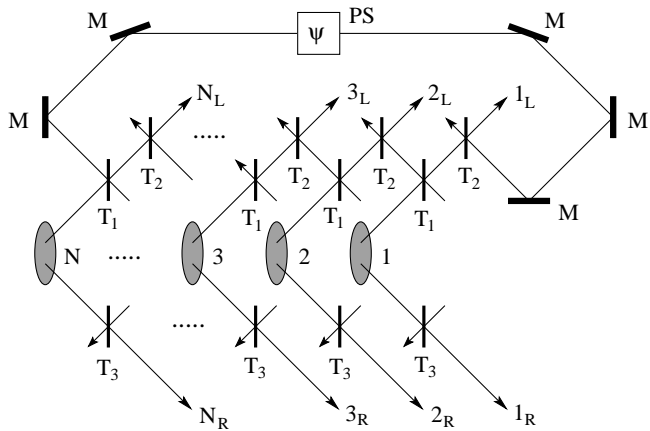,width=0.99\linewidth}}
\caption{Setup for the linear-optics $N$-qubit controlled phase gate. $T_j$ label
beam splitters with transmittance $T_j$, $M$ indicate mirrors, PS is a phase shift
by amount $\psi$. Each grey ellipse represents a single qubit carried by a single
photon propagating in two different spatial modes.}
\end{figure}

\section{Quantum Toffoli gate with linear optics}

In this section we will present and analyze the scheme which realizes quantum
$N$-qubit Toffoli gate in the coincidence basis. More precisely, the scheme 
conditionally applies the $N$-qubit controlled phase gate to the input qubits, 
whereby the  phase changes by $\pi$ if all qubits are in the 
state $|\tilde{1}\rangle$ and does not change otherwise,
\begin{equation}
U_{CP}^{\pi} |j_1,j_2,\ldots,j_N\rangle = e^{i\pi \prod_{k=1}^N j_k}
|j_1,j_2,\ldots,j_N\rangle,
\label{UCP}
\end{equation}
where $j_k \in \{\tilde{0},\tilde{1}\}$ and  $k=1,\ldots,N$. Note that the N-qubit  
controlled-phase (C-phase) gate and the Toffoli gate are equivalent 
up to single-qubit Hadamard transformations $H$ on the $N$th qubit (the target), 
$U_T=H_N U_{CP} H_N$. 

\subsection{Quantum optical C-phase gate}

The proposed optical setup is schematically sketched in Fig. 1. 
The qubits are encoded into states of single photons. We assume  the dual-rail
encoding where the two logical levels  $|\tilde{0}\rangle_j$ and $|\tilde{1}\rangle_j$ 
correspond  to two paths  $j_L$ and $j_R$ taken by a photon, 
$|\tilde{0}\rangle_j=|01\rangle_{j_Lj_R}$ and 
$|\tilde{1}\rangle_j=|10\rangle_{j_Lj_R}$. 
Note that other encodings such as polarization or time-bin are also possible and can be mutually
converted into each other by means of polarizing beam splitters and
unbalanced interferometers.

The operation of the C-phase gate  requires that the phase of the $N$-photon
state changes by  $\pi$ if and only if all photons are in the $L$ modes. In the
proposed scheme this is achieved by the $N$-photon interference on an array of
unbalanced beam splitters, see Fig. 1. First, each photon propagating in the 
$L$ mode is split into two modes on a beam splitter with intensity transmittance
$T_1$. Then pairs of beams originating from modes $j_L$ and $(j+1)_L$ interfere
on an array of $N$ beam splitters with transmittance $T_2$. Each  mode $j_R$ passes
through a beam splitter with transmittance $T_3$ which acts as a filter
balancing the amplitudes of the modes $j_L$ and $j_R$ after the application of the
gate.

The gate operates in the coincidence basis, i.e. it succeeds if a single photon
is detected in each pair of the output modes $j_L$ and $j_R$. Assume first that
at least one photon is in mode $k_R$. It is easy to see that in this case
the only way how the photons in the $j_L$ modes can reach the appropriate output
ports of the beam splitters is that each photon is transmitted through both beam
splitters and the total probability amplitude of this to happen 
reads $a_n=(t_1t_2)^{N-n}t_3^n$,
where $n \geq 1$ is the number of photons in $k_R$ modes and $t_j$ is the
amplitude transmittance, $T_j=t_j^2$. Since the gate should be
unitary, $a_n$ must not depend on $n$ which can be achieved by choosing 
\begin{equation}
T_3=T_1 T_2.
\end{equation}

The situation changes when all photons are initially in $L$ modes, i.e. the
input state reads $|\tilde{1},\tilde{1},\tilde{1},\ldots,\tilde{1}\rangle$. 
In this case, there are two  ways
how the photons can reach the $N$ output ports. One option is that all photons
are transmitted through all beam splitters. The second option is is 
that all photons are reflected from all beam splitters. Provided that these two 
alternatives are indistinguishable, i.e. there is a good spatiotemporal overlap
of the photonic wavepackets on the beam splitters, they interfere, and the
resulting amplitude reads,
\begin{equation}
a_0=t_1^Nt_2^N-r_1^N r_2^N.
\label{a0}
\end{equation}
Note the minus sign which arises due to the $\pi$ phase shift $\psi=\pi$ 
in one path of the reflected photon, see Fig. 1.
Note also that in the figure the path of the
reflected photon in mode $N_L$ looks much longer than all other paths. In the
actual implementation the geometry of the setup should be such that
all paths would be carefully balanced resulting in a good overlap of the photons 
and high-visibility interference. 

The gate operates as desired if $a_0=-a_{n > 0} =-t_1^N t_2^N$. Expressed in terms of the 
intensity transmittances this condition translates into
\begin{equation}
4 T_1^N T_2^N= (1-T_1)^N(1-T_2)^N,
\label{T12condition}
\end{equation}
where we used that $r_j=\sqrt{1-T_j}$. The formula  (\ref{T12condition}) describes a single-parametric
class of the N-qubit optical controlled-phase gates working in the coincidence
basis. The probability of success is given by 
$P_{\mathrm{succ}}=|a_n|^2=T_1^N T_2^N$ and
on expressing $T_2$ in terms of $T_1$ from Eq. (\ref{T12condition}),
\begin{equation}
T_2= \frac{1-T_1}{1-T_1 +4^{1/N} T_1},
\label{T2}
\end{equation}
we obtain
\begin{equation}
P_{\mathrm{succ}}=\left[\frac{T_1(1-T_1)}{1-T_1 +4^{1/N} T_1}\right]^N.
\label{Psucc}
\end{equation}
The optimal $T_1$ maximizing $P_{\mathrm{succ}}$ can be easily determined by solving
$d P_{\mathrm{succ}}/d T_1=0$, which yields
\begin{equation}
T_{1,\mathrm{opt}}=\frac{1}{1+2^{1/N}}.
\label{T1opt}
\end{equation}
On inserting this value into Eq. (\ref{T2}) we find that 
$T_{2,\mathrm{opt}}=T_{1,\mathrm{opt}}$ hence it
is optimal to use a scheme where the transmittances $T_1$ and $T_2$ are the
same. The optimal probability then reads,
\begin{equation}
P_{\mathrm{succ,opt}}= \left(\frac{1}{1+2^{1/N}}\right)^{2N}.
\label{Psuccopt}
\end{equation}
In particular, for $N=3$ we find $P_{\mathrm{succ,opt}}^{N=3}\approx 0.75$\%.

\subsection{Generalized C-phase gate}

The transformation  (\ref{UCP}) can be extended such that an arbitrary 
phase shift $\phi$ is introduced when all qubits are in logical state
$|\tilde{1}\rangle$. The generalized controlled-phase gate thus acts as follows,
\begin{equation}
U_{CP}^{\phi} |j_1,j_2,\ldots,j_N\rangle = e^{i\phi \prod_{k=1}^N j_k}
|j_1,j_2,\ldots,j_N\rangle.
\label{UCPtheta}
\end{equation}
We shall show that also this operation can be conditionally implemented with the
scheme shown in Fig. 1 provided that the phase shift $\psi$ in one arm of the multiphoton
interferometer and the transmittances $T_1$ and $T_2$ are properly chosen.
Repeating the derivation outlined in the preceding subsection we find that 
the condition that has to be satisfied reads
\begin{equation}
 e^{i\phi} t_1^Nt_2^N=t_1^N t_2^N+e^{i\psi} r_1^N r_2^N. 
 \label{trtheta}
\end{equation}
Upon splitting this formula into the real and imaginary parts and solving 
for $\psi$ we obtain
\begin{equation}
\tan \psi=-\frac{1}{\tan \frac{\phi}{2}}.
\label{psi}
\end{equation}
Similarly we also arrive at a generalization of the formula (\ref{T12condition}),
\begin{equation}
4 \, T_1^N \,T_2^N \sin^2\frac{\phi}{2} =(1-T_1)^N (1-T_2)^N.
\label{T1T2theta}
\end{equation}
The probability of success is maximized by choosing 
\begin{equation}
T_1=T_2=\frac{1}{1+|2\sin\frac{\phi}{2}|^{1/N}},
\label{Topttheta}
\end{equation}
and we have $P_{\mathrm{succ,opt}}(\phi)=1/[1+|2\sin(\phi/2)|^{1/N}]^{2N}$.
Note that the probability of success depends on the required conditional phase
shift $\phi$ and for a fixed $N$ it achieves its minimum for $\phi=\pi$, i.e.
when we attempt to implement the N-qubit Toffoli gate.

\begin{figure}

\centerline{\psfig{figure=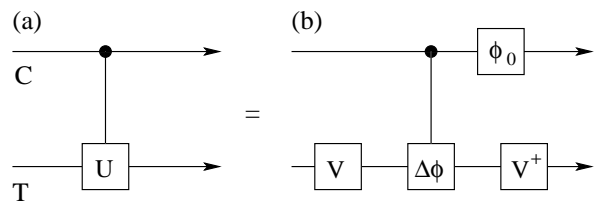,width=0.9\linewidth}}

\caption{Equivalence between two-qubit controlled unitary and controlled phase
gates.}

\end{figure}

With the two-qubit generalized C-phase gate at hand we can implement in the
coincidence basis an arbitrary two-qubit
controlled-$U$ gate, where a unitary operation $U$ is applied to the target
qubit iff the control qubit is in state $|\tilde{1}\rangle_C$. 
The equivalent scheme involving C-phase gate is shown in Fig. 2(b).
Note first that in the basis of eigenstates $|u_j\rangle_T$ of $U$, the controlled 
unitary gate boils down to conditional phase shifts, 
\begin{equation}
|\tilde{0}\rangle_C|u_j\rangle_T \rightarrow |\tilde{0}\rangle_C|u_j\rangle_T, 
\qquad
|\tilde{1}\rangle_C|u_j\rangle_T \rightarrow e^{i\phi_j}|\tilde{1}\rangle_C|u_j\rangle_T. 
\label{controlledU}
\end{equation}
 The unitary
$V$ maps the eigenstates of $U$ onto the computational basis states,
$V|u_j\rangle_T=|j\rangle_T$, $j=\tilde{0},\tilde{1}$. Next a C-phase gate with
$\Delta\phi=\phi_1-\phi_0$ follows. Finally, the inverse operation $V^\dagger$  
is applied to the target while the control qubit is subject to a phase shift operation
$|\tilde{0}\rangle_C\rightarrow |\tilde{0}\rangle_C$, $|\tilde{1}\rangle_C \rightarrow
e^{i\phi_0}|\tilde{1}\rangle_C$. It is easy to see that the net result of this sequence
of gates is the controlled-$U$ operation (\ref{controlledU}).

\subsection{Heralded controlled-phase gate}

The advantage  of working in the coincidence basis is that no extra ancillary
photons are required. The scheme is thus very economical in resources and for
instance the demonstration of the three-qubit Toffoli gate 
would require  detection of three-photon coincidences which is well within
the scope of present technology.

However, this approach also suffers from a significant disadvantage, since we 
do not know whether the gate succeeded until we detect the photons.
It is thus not possible to directly employ this gate as a part of a more
complex quantum information processing network. Nevertheless, it is possible to 
remove this drawback by performing quantum non-demolition measurements of
the number of photons at the outputs of the gate. If this measurement verifies that a
single photon is present in each pair of modes $j_L$ and $j_R$ then we
know that the gate was applied successfully while the non-demolition character
of this measurement guarantees that the output photons emerging from the gate
are preserved and not destroyed by the verification. 

A simple way of performing the non-demolition measurement of a number of
photons in two modes is to employ an auxiliary pair of photons in a maximally
entangled state and attempt to teleport the single photon in modes $j_L$ and
$j_R$ \cite{Kok02}. The single-photon detectors used for the partial Bell 
measurement which lies at the heart of the teleportation \cite{Bouwmeester97} 
must be able to resolve the number of photons. Detection of exactly two photons 
in the Bell analysis confirms that a single photon has been successfully
teleported. Observation of any other total number of photons indicates a
failure of the gate. This method requires $N$ auxiliary maximally entangled
photon pairs in total and the probability of successful non-demolition
measurement  given that the gate was applied successfully scales as $1/2^N$
because the optimal partial Bell measurement with linear optics can distinguish
only  two out of  four Bell states.

An alternative scheme for partial probabilistic non-demolition photon number measurement on a
pair of modes has been proposed in \cite{Kok02} (see also discussion 
in Ref. \cite{Fiurasek03}). The advantage of this latter scheme is that it does 
not rely on maximally entangled photon pairs and instead requires  single photons in
product state, which may be easier to generate.  The measurement requires two
ancilla photons and two photodetectors which can distinguish the number of
photons in a mode. A coincidence detection of a single photon by each detector 
indicates that at least a single photon has been present in the input pair of
modes and if exactly a single photon was at the input then its state was not
disturbed by the measurement.  If this partial measurement is carried out on
each pair of modes $j_L,~j_R$ and if all $N$ measurements indicate that there was at least a
single photon in each pair of modes then since there were altogether $N$ photons 
at the input of the gate we can conclude that the C-phase gate 
was applied successfully.

\begin{figure}[!t!]

\centerline{\psfig{figure=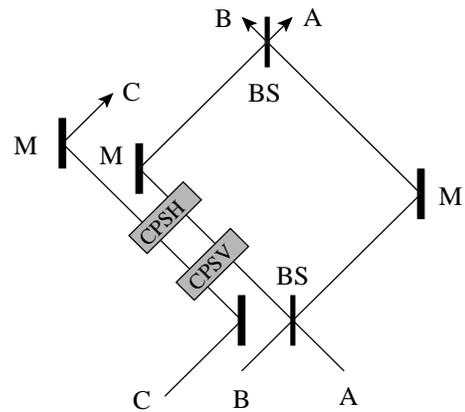,width=0.7\linewidth}}

\caption{Quantum optical Fredkin gate based on a balanced Mach-Zehnder interferometer. BS
denote balanced beam splitters and M indicate mirrors.  If the control
photon in mode C is vertically polarized then the boxes CPSV and CPSH
apply conditional phase shift $\pi$ to the  vertically (CPSV) or 
horizontally (CPSH) polarized modes in the left arm of the interferometer.} 

\end{figure}

\section{Fredkin gate}

Our scheme for linear optics Fredkin gate is inspired by the quantum optical
Fredkin gate originally proposed by Milburn \cite{Milburn89}. 
The Fredkin gate is a controlled
SWAP operating on the Hilbert space of three qubits, the states of qubits $A$
and $B$ are exchanged if the control qubit $C$ is in state $|\tilde{1}\rangle$ and nothing
happens if it is in state $|\tilde{0}\rangle$. Let us assume that the qubits are 
encoded onto polarization states of single photons. The controlled SWAP operation 
can be  converted to the controlled phase shift with the use of 
a balanced Mach-Zehnder interferometer, see Fig. 3. Depending on the state of 
the control photon $C$, the phase shift in the left arm of the
interferometer should be either $0$ or $\pi$, the latter results in the
effective swap of the photons $A$ and $B$ at the output of the interferometer.
In Milburn's scheme, the controlled phase shift is achieved by  medium 
with cross Kerr nonlinearity.

In the spirit of linear optics quantum computing, we suggest to replace the Kerr
medium with a linear interferometric scheme, and employ ancilla photons and
postselection conditioned on single photon detection to simulate the required
cross Kerr interaction \cite{Clausen03}. Since we assume polarization encoding, the conditional phase shift has to
be applied to both vertically and horizontally polarized modes in the left arm of
the Mach-Zehnder interferometer in Fig. 3. In what follows we will describe a
scheme which provides this conditional phase shift for a  single mode, and the
linear optics Fredkin gate then involves two such basic blocks acting in series
on vertically and horizontally polarized mode, see Fig. 3.

\begin{figure}[!t!]

\centerline{\psfig{figure=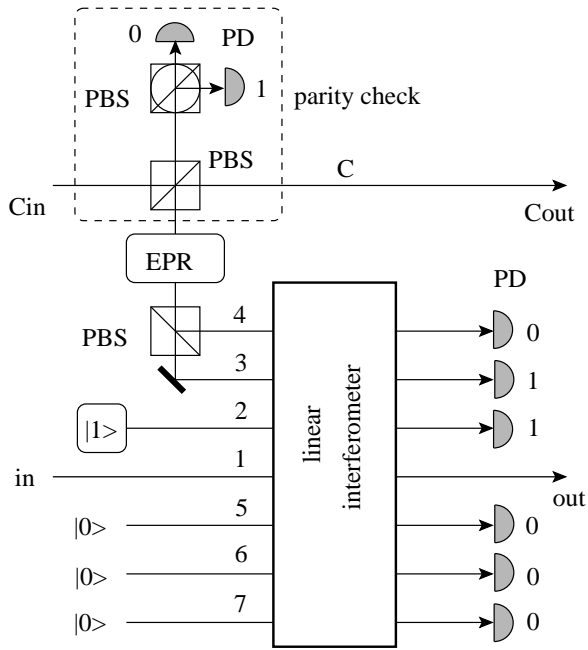,width=0.9\linewidth}}

\caption{Simulation of cross-Kerr interaction with linear optics. 
The setup involves polarizing beam splitters (PBS),
source of an auxiliary entangled pair of photons (EPR), 
source of a single photon ($|1\rangle$), linear multiport interferometer 
consisting of beam splitters and phase shifters and single-photon detectors (PD).} 

\end{figure}

The basic block is depicted in detail in Fig. 4. The scheme requires three
ancilla photons: a maximally entangled pair of photons in a state 
$\frac{1}{\sqrt{2}}(|VV\rangle+|HH\rangle)$ emitted by source (EPR) and an additional single photon in
mode $2$. The proposed setup consists of two main parts. The first part 
is the quantum parity check \cite{Pittman01,Pittman02}
between the control photon in mode $C$ and one photon from the
auxiliary EPR beam. The check is based on a coupling of these photons on a
polarizing beam splitter PBS followed by a detection of one of the outputs in
the basis $\frac{1}{\sqrt{2}}(|V\rangle \pm |H\rangle)$. The detectors should be able to resolve the
number of photons in the beam and the parity check is successful if a single
photon is detected by one detector and no photon is detected by the other
detector, which happens with probability $1/2$. 
The parity check effectively copies in the computational basis the state of 
the control photon in spatial mode $C$ onto the auxiliary photon in spatial 
modes $3$ and $4$ and we can write,
\begin{equation}
\alpha|V\rangle_C+\beta|H\rangle_C \rightarrow 
\alpha|V\rangle_C|0\rangle_3|1\rangle_4+\beta|H\rangle_C|1\rangle_3|0\rangle_4 .
\label{paritycheck}
\end{equation}
The parity check allows us to control the phase shift of mode $1$ indirectly by
the auxiliary photon while preserving the original control photon. A  similar
trick has been used in the recent experimental implementations of quantum C-NOT
gate \cite{Pittman03,Gasparoni04,Zhao05}.

The second part of the scheme in Fig. 4 consists of a linear interferometer
where the photons in the mode $1$ are combined with the ancilla photons in modes
$2$, $3$ and $4$. Note that the interferometer has also three other auxiliary
input ports in vacuum state. All output modes of the interferometer except 
for mode $1$ are monitored with photon number resolving detectors and the 
conditional phase shift is successfully applied if a single photon 
is detected in modes $2$ and $3$ and no photons are observed in the other modes. 

The purpose of the interferometer is to conditionally induce a phase shift $\pi$
in mode $1$ provided that there is a photon in the input mode $4$ and induce no
shift if the photon is in mode $3$. Since there can be no more than $2$ photons
in the mode $1$ (the two photons whose polarization states should be
conditionally swapped), it suffices to achieve the correct conditional phase
shift in the subspace of Fock states $|0\rangle_1$, $|1\rangle_1$ and
$|2\rangle_1$.

Mathematically, the interferometer is described by a unitary matrix $U$,
which governs the transformation between input and output modes. 
The input creation operators $a_{\mathrm{in},j}^\dagger$ can be expressed as linear
superpositions of the output creation operators $a_{\mathrm{out},k}^\dagger$ 
according to
\begin{eqnarray}
a_{\mathrm{in},j}^\dagger=\sum_{k=1}^7 u_{jk} a_{\mathrm{out},k}^\dagger ,
\label{interferometer}
\end{eqnarray}
where $u_{jk}$ are the elements of $U$. Since 
we condition on observing no photons in  modes $4$-$7$, in our subsequent 
calculations we will explicitly  need only the coefficients $u_{jk}$ with
$j=1,2,3,4$ and $k=1,2,3$.

Due to the linearity we can treat separately the cases when the control photon
is in state $|H\rangle$ and $|V\rangle$. Assume first that it is in state
$|H\rangle$. The two auxiliary photons are then in input modes $2$ and $3$ and 
conditionally on detecting a single photon in the output modes $2$ and $3$
and no photon is all other modes (except for mode $1$ which is not measured upon),
we obtain the following transformation,
\begin{widetext}
\begin{eqnarray}
|H\rangle_C|1\rangle_2|1\rangle_3|0\rangle_4
(\alpha_0|0\rangle_1+\alpha_1|1\rangle_1+\alpha_2|2\rangle_1) \rightarrow
|H\rangle_C(x_0 \alpha_0 |0\rangle_1+x_1\alpha_1|1\rangle_1+x_2 \alpha_2|2\rangle_1) 
\label{conditionalH}
\end{eqnarray}
where the coefficients $x_j$ can be expressed as follows,
\begin{eqnarray}
x_0&=&u_{22}u_{33}+u_{23}u_{32}, \nonumber \\
x_1&=&u_{11}(u_{22}u_{33}+u_{23}u_{32})+u_{12}(u_{23}u_{31}+u_{21}u_{33})+
u_{13}(u_{22}u_{31}+u_{21}u_{32}), \nonumber \\
x_2&=&u_{11}^2(u_{33}u_{22}+u_{32}u_{23})+2u_{12}u_{13}u_{21}u_{31}+
2u_{11}u_{12}(u_{23}u_{31}+u_{21}u_{33})+2u_{11}u_{13}(u_{22}u_{31}+u_{21}u_{32}).
\label{x}
\end{eqnarray}
If the control photon is in state $|V\rangle$, then the conditional
transformation reads
\begin{eqnarray}
|V\rangle_C|1\rangle_2|0\rangle_3|1\rangle_4
(\alpha_0|0\rangle_1+\alpha_1|1\rangle_1+\alpha_2|2\rangle_1) \rightarrow
|V\rangle_C(y_0 \alpha_0 |0\rangle_1+y_1\alpha_1|1\rangle_1+y_2
\alpha_2|2\rangle_1) ,
\end{eqnarray}
\label{conditionalV}
 \end{widetext}
where the coefficients $y_j$ can be expressed in the same way as $x_j$,
only the matrix elements $u_{3k}$ in Eq. (\ref{x}) must be replaced with $u_{4k}$.

We want to implement a conditional $\pi$-phase shift in mode $1$. This will be
achieved if  
\begin{equation}
x_j=q, \qquad y_j=q(-1)^j,
\label{xycondition}
\end{equation}
where $q<1$ is some shrinking factor arising due to the probabilistic nature of
the gate.  Low $q$ reduces the probability of success of the gate 
which scales as $P\propto|q|^2$ but does not alter its operation. 
The maximum $q$ that can be chosen is
determined by the constraint that $u_{jk}$ ($1 \leq j \leq 4 $, $1 \leq k \leq 3$)
must form a submatrix of a unitary
matrix.  As shown in the Appendix, it is possible to efficiently numerically 
determine whether a given set of $u_{jk}$ may form a submatrix 
of $U$ so that also the maximum $q$ can be determined numerically.

By solving the system of nonlinear equations (\ref{xycondition}) 
we can find the matrix elements $u_{jk}$ specifying the interferometer 
which implements the conditional phase shift. Note that the system 
is underdetermined hence there exist infinitely many interferometers 
satisfying (\ref{xycondition}). To see this, it is convenient to rewrite 
these equations in a matrix form,
\begin{equation}
\bm{M} \bm{u}_3= \left(
\begin{array}{r} q \\ q \\ q \end{array}\right),
 \qquad \bm{M} \bm{u}_4= \left(
\begin{array}{r} q \\ -q \\ q \end{array}\right),
\label{Matrixform}
\end{equation}
where $\bm{u}_3=(u_{31},u_{32},u_{33})^T$ and 
$\bm{u}_4=(u_{41},u_{42},u_{43})^T$ are
column vectors and $\bm{M}$ is a matrix whose elements can be expressed in terms
of $u_{1k}$ and $u_{2k}$. Thus when looking for the solution to Eqs.
(\ref{Matrixform})   we can
choose arbitrary $u_{1k}$ and $u_{2k}$ and provided that $\det \bm{M} \neq 0$
we can for a given $q$ calculate $\bm{u}_3$ and $\bm{u}_4$ by solving the system
of linear equations (\ref{Matrixform}). 

Let us now present a particular example of an analytical solution. Choosing
\begin{equation}
u_{11}=u_{12}=u_{13}, \qquad
u_{21}=-u_{22}=u_{23}, 
\label{uaa}
\end{equation}
we can express the matrix elements $u_{3k}$  as follows,
\begin{eqnarray}
u_{31}&=&-\frac{q}{2u_{11}^2 u_{22}}(1-u_{11})^2, \nonumber \\
u_{32}&=&-\frac{q}{2u_{11}u_{22}}, \nonumber \\
u_{33}&=&\frac{q}{2 u_{11}u_{22}}(2u_{11}-1). 
\label{uck}
\end{eqnarray}
Similar formulas hold also for $u_{4k}$ and we have,
\begin{eqnarray}
u_{41}&=&-\frac{q}{2 u_{22} u_{11}^2 }(1+u_{11})^2, \nonumber \\
u_{42}&=&\frac{q}{2u_{11}u_{22}}, \nonumber \\
u_{43}&=&\frac{q}{2 u_{11}u_{22}}(2u_{11}+1). 
\label{udk}
\end{eqnarray}
The maximum $|q|$ achievable within the above given analytical solution  was 
determined numerically and we found that it is optimum to choose 
$u_{11,\mathrm{opt}}=0.494$ and $u_{22,\mathrm{opt}}=0.416$ 
yielding $q_{\mathrm{opt}}=0.0638$. Since the Fredkin gate in Fig. 3 includes
two conditional phase shift gates, the total probability of success of the gate
is given by $P_{\mathrm{succ}}=\frac{1}{4}|q|^4$ where the factor $\frac{1}{4}$ 
appears due to the
two quantum parity checks. On inserting the $q_{\mathrm{opt}}$ into this formula we
obtain $P_{\mathrm{succ}}\approx 4.2 \times 10^{-6}$, which is rather small. 
However, it should be stressed that this is not the maximum probability of
success that could be attained with our scheme. It is possible to improve the
success rate by several orders of magnitude  by performing numerical optimization over 
all relevant parameters $u_{1k}$  and $u_{2k}$. We have carried out 
a thorough numerical search and the maximum $P_{\mathrm{succ}}$ that we obtained in
this way reads $P_{\mathrm{succ,max}}=4.1 \times 10^{-3}$. 

In is instructive to compare this value with the probability of success that
could be achieved if one would attempt to implement the Fredkin gate as a sequence
of two-qubit unitaries. It was shown by Smolin and DiVincenzo that five 
two-qubit quantum gates suffice to implement the Fredkin gate \cite{Smolin96}. 
Making the very optimistic assumption that using two ancilla photons per gate 
each of these gates  can be implemented with probability $1/4$ 
similarly as the C-NOT \cite{Gasparoni04,Zhao05} we arrive at 
a total probability  $P_{\mathrm{succ}}^\prime=4^{-5}\approx 9.8 \times 10^{-4}$.
Thus our scheme for Fredkin gate could potentially attain a higher probability
of success while being more economical in resources because it
requires only $6$ ancilla photons instead of $10$ photons.

\section{Conclusions}

We have devised schemes for linear optics quantum Toffoli and Fredkin gates. 
In the spirit of linear optics quantum computing the gates do not require
nonlinear interaction and instead rely on multiphoton interference, ancilla
photons and postselection conditioned on single-photon detection. 
The key feature of the proposed setups is that they are directly tailored
for the implementation of the  multiqubit Toffoli or Fredkin gate. This should
be contrasted with the common approaches 
where the multiqubit gates are decomposed into a sequence of two- and
single-qubit gates. 

Given the current state of the technology, our direct
approach to multiqubit gates may be much more efficient than implementations
relying on a sequence of two-qubit gates. In particular, the experimental
demonstration of the three-qubit quantum Toffoli gate  in the coincidence basis
would require only three photons and an observation of a three-photon
coincidences, which is well within the reach of current technology.

Despite their advantages, the present schemes still suffer from some weaknesses. 
The probability of success of the N-qubit Toffoli gate exponentially decreases
with growing $N$ and also the probability of success of the Fredkin gate,
$P_{\mathrm{succ}}\approx 4.1 \times 10^{-3}$, is not very high. Another drawback lies in
the fact that setups for both Toffoli and Fredkin gate require interferometric 
stability, which is hard to achieve and maintain. In contrast, recent
experimental demonstrations of the quantum linear-optical C-NOT gate relied
solely on Hong-Ou-Mandel interference effect \cite{Langford05,Kiesel05,Okamoto05}, 
which is much more robust against small length fluctuations.

It remains an interesting open question whether a scheme avoiding problems with
interferometric stability could be devised also for the Toffoli and Fredkin
gates. Another important open issue is what is the maximum achievable success
rate for these gates either without ancillas (i.e. operating in the coincidence
basis) or with a given fixed amount of auxiliary photons. We hope that the
present paper will stimulate further  theoretical as well as experimental
investigations along these lines potentially resulting in an important step
towards linear optics quantum computing.

\begin{acknowledgments}
The author would like to thank L. Mi\v{s}ta for helpful discussions and comments.
This work was supported under the Research 
project Measurement and Information in Optics MSM 6198959213 
of the Czech Ministry of Education.

\end{acknowledgments}

\appendix*

\section{Testing the unitarity}

Here we show how to test whether a $4\times 3$ matrix $u_{jk}$, 
$1\leq j \leq 4$,  $1\leq k \leq 3$, could be a submatrix of a larger $7\times 7$ 
unitary matrix $U$. The procedure consists of two steps which have to be repeated 
for each $j\in\{1,2,3,4\}$. We start from $j=1$. 

Step (i): We set $u_{jk}=0$ for $k> j+3$. We can assume without loss of generality
that $u_{j,j+3}$ is real and nonnegative and determine it from the normalization
condition $\sum_k |u_{jk}|^2=1$,
\[
u_{j,j+3}= \sqrt{1-\sum_{k=1}^{j+2}|u_{jk}|^2}.
\]
If this expression yields purely imaginary $u_{j,j+3}$ then $u_{jk}$ could not 
form a part of a unitary matrix and we terminate the test.

Step (ii): We must guarantee that the rows of $U$ are mutually orthogonal. This can
be achieved  by calculating $u_{k,j+3}$ from the orthogonality condition
\[
\sum_l u_{jl}^\ast u_{kl} = 0, \qquad j \neq k.
\]
For all $k$ satisfying $j<k \leq 4$ we thus have
\[
u_{k,j+3}= -\frac{1}{u_{j,j+3}^\ast} \sum_{l=1}^{j+2} u_{jl}^\ast u_{kl}.
\]
Finally, we increase $j$ by $1$.

The steps (i) and (ii) have to be repeated until $j=4$ is reached. If all four
iterations succeed then at the end we obtain a $4 \times 7$ isometry matrix that
could be easily completed to form a unitary matrix. Otherwise we know from step (i) 
that the matrix $u_{jk}$ could not form a submatrix of a unitary matrix.

\end{document}